\documentclass[review, 11pt]{elsarticle}

\usepackage{amssymb}
\usepackage{amsmath}
\usepackage{listings}
\usepackage[export]{adjustbox}

\usepackage[]{xcolor}
\usepackage{siunitx}
\usepackage{lineno}
\usepackage{multirow}
\usepackage{textgreek}
\usepackage{mathtools}
\usepackage[capitalize]{cleveref}
\usepackage{hyphenat}

\usepackage{subcaption}
\usepackage{pgfplots}
\usepackage{booktabs} 
\pgfplotsset{compat=1.18}
\usepackage{tikz}
\usetikzlibrary{fit} 
\usepackage{float}
\usepackage{threeparttable}

\usepackage{relsize}

\usepackage{xcolor}
\definecolor{TUMBlau}{RGB}{0,100,189} 
\definecolor{TUMBlauDunkel}{RGB}{0,82,147} 
\definecolor{TUMBlauSehrDunkel}{RGB}{0,51,81} 
\definecolor{TUMBlauHell}{RGB}{152,198,234} 
\definecolor{TUMBlauMittel}{RGB}{100,160,200} 
\definecolor{TUMElfenbein}{RGB}{218,215,203} 
\definecolor{TUMGruen}{RGB}{162,173,0} 
\definecolor{TUMOrange}{RGB}{227,114,34} 
\definecolor{TUMGrauZwanzig}{RGB}{217,218,219}

\usepackage{pgfplots}
\pgfplotsset{compat=newest}

\usepgfplotslibrary{colormaps}
\usepgfplotslibrary{fillbetween}
\usetikzlibrary{arrows.meta}

\pgfplotscreateplotcyclelist{TUMcyclelist}{%
	TUMBlau,every mark/.append style={solid,fill=TUMBlau},mark=*\\%
	TUMOrange,every mark/.append style={solid,fill=TUMOrange},mark=diamond*\\%
	TUMGruen,every mark/.append style={solid,TUMGruen},mark=x\\%
	TUMBlauDunkel,every mark/.append style={solid,fill=TUMBlauDunkel},mark=triangle*\\%
	TUMBlauMittel,every mark/.append style={solid,fill=TUMBlauMittel},mark=diamond*\\%
	TUMBlau,densely dashed,every mark/.append style={solid,fill=TUMBlau},mark=triangle*\\%
	TUMOrange,densely dashed,every mark/.append style={solid,fill=TUMOrange},mark=square*\\%
	TUMBlauDunkel,densely dashed,mark=star,every mark/.append style={solid,fill=TUMBlauDunkel},mark=x\\%
	TUMBlauMittel,densely dashed,every mark/.append style={solid,fill=TUMBlauMittel},mark=square*\\%
	TUMGruen,densely dashed,every mark/.append style={solid,TUMGruen},mark=*\\%
}

\pgfplotscreateplotcyclelist{TUMcyclelistNomark}{%
	TUMBlau,solid,mark=none\\
	TUMOrange,dashed,mark=none\\
	TUMGruen,dotted,mark=none\\
	TUMBlauDunkel,dashdotted,mark=none\\
	TUMBlauHell,densely dashed,mark=none\\
	TUMBlau,densely dotted,mark=none\\
	TUMOrange,densely dashedotted,mark=none\\
	TUMGruen,loosely dashed,mark=none\\
	TUMBlauDunkel,loosely dotted,mark=none\\
	TUMBlauHell,loosely dashedotted,mark=none\\
}

\usetikzlibrary{patterns}

\usetikzlibrary{shapes}
\tikzset{cross/.style={cross out, draw,
			minimum size=2*(#1-\pgflinewidth),
			inner sep=0pt, outer sep=0pt}}

\tikzset{itneighbor/.style={fill=black,regular polygon, regular polygon sides=4,inner sep=2.5}}
\tikzset{itextension/.style={fill=black,regular polygon, regular polygon sides=3,inner sep=2.0}}
\tikzset{itreinit/.style={fill=black, circle, inner sep=2.5}}
\tikzset{itcutcell/.style={fill=black,cross,inner sep=3.5,very thick}}

\usepackage[nolist]{acronym}

\begin{acronym}

    \acro{fno}[FNO]{Fourier Neural Operator}
    \acro{lbm}[LBM]{Lattice Boltzmann Method}
    \acro{nse}[NSE]{Navier-Stokes Equations}
    \acro{pcc}[PCC]{Pearson Correlation Coefficient}
    \acro{mse}[MSE]{Mean Squared Error}
    \acro{pde}[PDE]{Partial Differential Equation}
    \acro{fem}[FEM]{Finite Element Method}
    \acro{srt}[SRT]{Single Relaxation Time}
    \acro{l}[L]{Layers}
    \acro{lc}[LC]{Latent Channels}
\end{acronym}

\journal{Elsevier}

\begin{document}

\begin{frontmatter}

	\title{Hybrid Fourier Neural Operator-Lattice Boltzmann Method}
	\author[tumaer]{A.~Junk\corref{cor1}}
	\ead{alexandra.junk@tum.de}

	\author[tumaer]{J.~M.~Winter}
	\ead{josef.winter@tum.de}

    \author[tumaer, addr1]{M.~Tütken}
	\ead{meike.tuetken@siemens.com}
    
	\author[tumaer]{S.~Schmidt}
	\ead{steffen.schmidt@tum.de}

	\author[tumaer]{N.~A.~Adams}
	\ead{nikolaus.adams@tum.de}

	\address[tumaer]{Technical University of Munich, Department Engineering Physics and Computation, Chair of Aerodynamics and Fluid Mechanics, Boltzmannstra\ss e 15, 85748 Garching, Germany}


    \cortext[cor1]{Corresponding author.}
    \fntext[addr1]{Present address: 
    Siemens Industry Software NV, Interleuvenlaan 68, 3001 Heverlee, Belgium; 
    KU Leuven, Department of Mechanical Engineering, Celestijnenlaan 300, 3001 Heverlee, Belgium}

\begin{abstract}
We propose an accelerated computational fluid dynamics framework based on a hybrid Fourier Neural Operator–Lattice Boltzmann Method (FNO–LBM) for steady and unsteady weakly compressible flows. FNO‑based initialization significantly accelerates LBM in reaching steady-states of porous media flows across all macroscopic fields, achieving up to 70\% speed‑up in convergence of density and more than 40\% of pressure‑drop while preserving the final steady‑state accuracy. Simulations of unsteady flows can be accelerated by hybrid coupling strategies that employ FNO rollouts embedded into LBM time advancement in a way of super-time-stepping. Global and time‑resolved error metrics across 100 trajectories for generic 2D flows demonstrate that hybridization consistently improves accuracy and stabilizes long‑horizon rollouts. Best efficiency is achieved for a lightweight 2.6M‑parameter FNO, which diverges under pure autoregressive rollout but achieves 96–99.8\% error reduction under hybrid coupling, matching the predictive capability of a much more expensive 11.2M‑parameter model. The hybrid framework enhances predictive fidelity, suppresses error accumulation, and enables small and cheap surrogate models to operate effectively within the same error regime as larger surrogates. These results demonstrate that hybrid neural‑operator coupling achieves robust and computationally efficient accelerated LBM while maintaining physically consistent flow evolution. 
\end{abstract}


\end{frontmatter}

\acresetall
\section{Introduction}
\label{sec:introduction}
\par High-fidelity numerical simulation of fluid flows remains central to fluid dynamics analysis, optimization, and design across science and engineering. However, the computational cost associated with resolving complex geometries, multiscale features, and long-time unsteady dynamics continues to pose significant challenges. The \ac{lbm} offers a robust, efficient and accurate computational framework for weakly compressible flows, but as with all classical CFD methods its cost scales with grid resolution and simulation time, limiting applications to problems that require rapid turnaround, extensive parameter studies, or real-time feedback \cite{Krüger2017}. These constraints have motivated growing interest in machine learning surrogate approaches capable of replacing and accelerating CFD workflows while preserving physical accuracy \cite{caron2025}.
\par With data-driven machine‑learning (ML) models for flow evolution, surrogate modeling has therefore emerged as a promising pathway for accelerating fluid simulations \cite{greenfeld2019learning, jiang2020meshfreeflownet, raissi2019nn, kochkov2021machine, li2021fourier}. However, many conventional neural network architectures operate on fixed spatial grids and struggle to generalize across discretizations or maintain stability during long autoregressive rollouts, restricting their reliability in high-resolution or long-horizon unsteady simulations. Neural operators address these limitations by learning mappings between function spaces rather than discrete arrays, enabling resolution-invariant inference and single-pass prediction of entire flow fields \cite{li2020neural, bhattacharya2021modelred, Lu2021, patel2021operator}. Among them, the \ac{fno} has emerged as a prominent architecture due to its spectral convolution formulation and quasi-linear scaling with resolution \cite{li2021fourier}. \ac{fno}s and their variants -- such as U-\ac{fno} \cite{Wen2022}, F-\ac{fno} \cite{tran2023}, S\ac{fno} \cite{bonev2023}, and multi-scale extensions \cite{YOU2026} -- have demonstrated impressive performance across a wide range of physics-informed tasks, including porous media flows \cite{choubineh2023}, elastodynamics \cite{LEHMANN2025}, hydraulic tomography \cite{Guo2024}, nonlinear optics \cite{MARGENBERG2024}, and tokamak plasma evolution \cite{Gopakumar2024}, establishing operator learning as a versatile tool for approximation of high‑dimensional PDE solutions.%
\par Despite their expressiveness, computationally cheap \ac{fno} models may rapidly accumulate error during autoregressive rollout and diverge after only a few autoregressive steps \cite{tran2023, bonev2023,Gopakumar2024}. This limitation has prompted increasing interest in hybrid modeling frameworks that combine machine learning surrogates with numerical solvers. By intermittently invoking directly a consistent approximation of the governing equations, hybrid approaches can correct accumulated errors, enforce physical constraints, and stabilize long-term predictions. Several such hybrid CFD frameworks have recently emerged. In the context of \ac{lbm}, Rabbani and Babaei \cite{RABBANI2019} proposed a machine-learning-accelerated permeability workflow, Zhang et al.\ \cite{Zhang2025} combined ML and \ac{lbm} for heat-flux prediction in phase-change processes, and Fischer et al. \cite{fischer2025MARL} used a reinforcement-learning closure model to stabilize under-resolved \ac{lbm} simulations. More broadly, hybrid ML–numerical solver systems include \ac{fno}-coupled finite-difference solvers for fluid–structure interaction \cite{xiao2024}, neural networks used as pressure solvers or closure models in Navier–Stokes flows \cite{SOUSA2024} and DeepONet preconditioning \cite{karniadakis2025}. These developments suggest that hybridization can combine the speed of ML surrogates with the robustness of physics-based solvers, but the degree to which hybrid frameworks can compensate for limited surrogate capacity or enforce long-horizon stability remains insufficiently explored \cite{caron2025}.
\par In this work, we address this gap by investigating hybrid integration of \ac{fno}s within an \ac{lbm} framework for both steady and unsteady weakly compressible flows. First, we employ \ac{fno}-predicted macroscopic fields for acceleration of steady-state \ac{lbm} predictions of porous media flows. Second, we consider a generic unsteady 2D shear flow, which is highly sensitive to accumulated errors and thus is a critical test of temporal surrogate stability and for assessing hybrid strategies in autoregressive rollout. In this case, the \ac{fno} is employed as a super-time-stepping algorithm that replaces a large number of \ac{lbm} time steps. We compare a large 11.2M-parameter \ac{fno} with a lightweight 2.6M-parameter model and find that numerical corrections are crucial for stabilization of the lightweight model during autoregressive rollout.
Our contributions are threefold:
\begin{itemize}
    \item \textbf{Steady-state-convergence acceleration:} \ac{fno}-initialized \ac{lbm} achieves substantial reductions in convergence time, reducing iteration counts by up to 70\% for density and over 40\% for pressure-drop metrics, with no loss of accuracy relative to converged steady-states obtained with \ac{lbm} only.
    \item \textbf{Improvement of autoregressive \ac{fno} rollouts by hybridization:} Hybrid \ac{fno}--\ac{lbm} coupling significantly improves accuracy even for autoregressivly stable \ac{fno} rollouts by up to $49.3\%$ and decreases total simulation runtimes by up to $11.8\%$.
    \item \textbf{Stabilization of low-cost \ac{fno} surrogates by hybridization:} Hybridization enables a 2.6M-parameter \ac{fno}, which is unstable as a standalone surrogate, to achieve 96\% to 99.8\% error reduction and reach the accuracy regime of the full 11.2M-parameter surrogate.
\end{itemize}
These results demonstrate that hybrid neural-operator frameworks offer a powerful and computationally efficient means of accelerating \ac{lbm} while maintaining physical fidelity and transforming compact surrogates into physically reliable models for long-horizon unsteady prediction. 

\section{Methodology}
\label{sec:methods}

\subsection{The Lattice Boltzmann Method}

\par All simulations are run in single-precision using TorchLBM \cite{winter2024}, a fully-differentiable and modular PyTorch implementation of an \ac{lbm} solver for the weakly compressible \ac{nse}. \ac{lbm} is a mesoscopic numerical method \cite{Chen1998} that computes the evolution of discrete particle distribution functions $f_i$ along predefined discrete lattice velocities $\bar{c}_i$ over a structured lattice grid. 
\begin{figure}[htbp]
    \centering
    \begin{subfigure}[c]{0.8\textwidth}
        \resizebox{\linewidth}{!}{\includegraphics[]{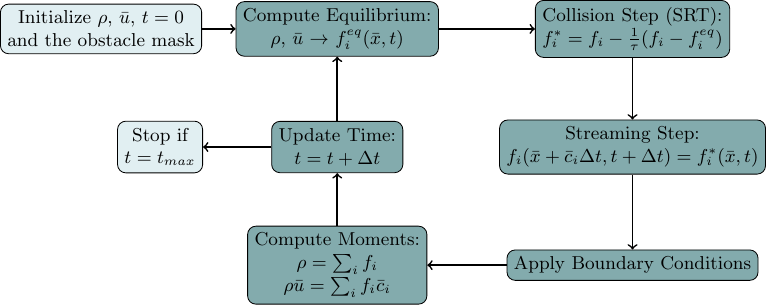}}
    \end{subfigure}
    \caption{Visualization of the \ac{lbm} simulation loop. If no macroscopic fields are available, they are set to default $\rho= \SI{1.0}{\frac{kg}{m^{3}}}$, $\bar{u} = \SI{0.0}{\frac{m}{s}}$ and no obstacle mask.}
    \label{fig:lbm_flowchart}
\end{figure}%
\par The general simulation loop of TorchLBM is depicted in \cref{fig:lbm_flowchart} and consists of six recurrent modules and one initialization block at the start of the simulation. After initialization of the macroscopic density $\rho$ and velocity $\bar u$, the main simulation loop starts with the calculation of the equilibrium distribution $f_i^{eq}$ following 
\begin{equation}
f_i^{eq}=w_i \rho \left( 1+\frac{\bar{u}\cdot\bar{c}_i}{c_s^2}+\frac{(\bar{u}\cdot\bar{c}_i)^2}{2c_s^4}-\frac{\bar{u}\cdot\bar{u}}{2c_s^2}\right),
\end{equation}
where $w_i$ denotes the respective lattice weights and $c_s^2$ the isothermal model's speed of sound. The number and configuration of lattice velocities and weights follow the $DdQq$ convention, with $d$ spatial dimensions and $q$ discrete velocities. For 2D configurations best efficiency is reached with the $D2Q9$ scheme. 
\par The equilibrium calculation is followed by the two split operations of the \ac{lbm} algorithm: collision and streaming. During the collision step, local non-linear particle interactions are modeled by the collision operator. For simplicity, we adopt here the \ac{srt} collision model
\begin{equation}
f_i^*(\bar{x},t) = f_i(\bar{x},t) \left( 1+\frac{\Delta t}{\tau} \right) + f_i^{eq}(\bar{x},t)\frac{\Delta t}{\tau}
\end{equation}
with the relaxation time $\tau$. In the streaming step, post-collision distributions $f_i^*$ are propagated through a linear, non-local streaming operation to neighboring nodes following
\begin{equation}
    f_i(\bar{x} + \bar{c}_i \Delta t, t+\Delta t) = f_i^*(\bar{x},t).
\end{equation}
Lastly, the macroscopic quantities are recovered from distribution moments
\begin{align}
    \rho(\bar{x},t) &= \sum_i f_i(\bar{x},t), \\
    \rho\bar{u}(\bar{x},t) &= \sum_i \bar{c}_i f_i(\bar{x},t),
\end{align}
and the simulation time is updated.
\par A comprehensive discussion of the \ac{lbm} and its numerical implementation is provided e.g. by Krüger et al. \cite{Krüger2017}, Succi \cite{succi2001} and Lallemand et al. \cite{Lallemand2021}.
\par The macroscopic density and velocity fields are employed as inputs for training and validation of the \ac{fno} models. Although \ac{lbm} naturally provides both mesoscopic and macroscopic data, employing macroscopic quantities enables transfer of the hybrid framework to other surrogate models trained on macroscopic data.
\subsection{Fourier Neural Operator}
\begin{figure}
    \centering
    \begin{subfigure}[c]{0.99\textwidth}
        \resizebox{\linewidth}{!}{\includegraphics[]{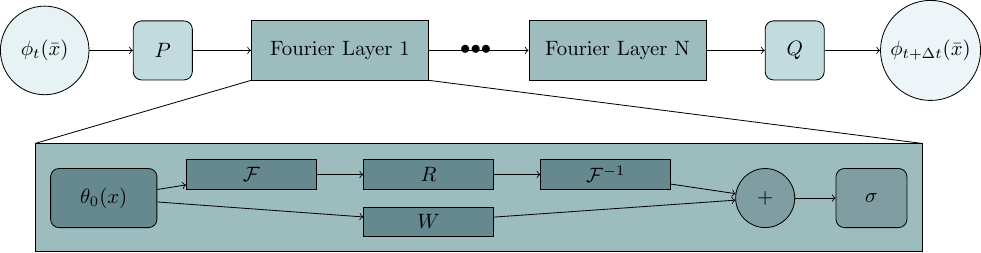}}
    \end{subfigure}
    \caption{FNO Architecture with the lifting network $P$, the Fourier transform $\mathcal{F}$, the learnable filter $R$, the local linear transform $W$, the activation $\sigma$ and the projection $Q$ back to physical space.}
    \label{fig:fnoarchitecture}
\end{figure}

Li et al. \cite{li2021fno} introduced the \ac{fno} as a discretization-invariant architecture that learns mappings between infinite-dimensional function spaces. The architecture builds on the Neural Operator framework proposed by Kovachki et al. \cite{li2021neural}, replacing the kernel integral operator with a global convolution in Fourier space, leading to the computationally efficient approximation.
The architectural structure of the \ac{fno} used in this study is illustrated in \cref{fig:fnoarchitecture}.
\par Let $\phi_t(\bar x) \in \mathbb{R}^{d}$ denote the input function, in our case the density and velocity fields $\phi_t(\bar x)=(u_t,v_t,\rho_t)$ at time $t$. This input function is first lifted into latent space by the shallow fully-connected neural network $P$ via $P(\phi(x))=\theta_0(x)$. The \ac{fno} then performs iterative updates $\theta_n \mapsto \theta_{n+1}$ for $n=0,1,...,N-1$ across the $N$ Fourier layers following 
\begin{equation}
\theta_{n+1}(x) = \sigma \Big( W \theta_n(x) + \mathcal{F}^{-1} \big( R(\bar{k}) \cdot \mathcal{F}(\theta_n)(\bar{k}) \big) \Big),
\end{equation}
where $\sigma$ is a pointwise nonlinearity (ReLU), $W$ is a local linear transformation, $\mathcal{F}$ and $\mathcal{F}^{-1}$ denote the Fourier and inverse Fourier transforms, $\bar{k}$ represents the cut-off filtered frequency modes, and $R$ is a learnable complex-valued filter kernel applied to each Fourier mode. 
\par Lastly, the output is projected back to physical space with $Q(\theta_{N-1})=\phi_{t+ \Delta t}(\bar x)$, where in our case $\Delta t = \Delta T_{FNO} = 223 \cdot \Delta t_{LBM}$ represents the discrete time-step size of the FNO. 
\subsection{Hybrid FNO-LBM Framework}
\label{sec:hybrid_framework}
We propose a hybrid framework that couples a fully trained \ac{fno} with the \ac{lbm} solver, thereby exploiting the complementary strengths of both approaches. Within this framework, the \ac{fno} is employed either as a generator of an initial flow field for steady-state problems or during time marching, functioning as a super-time-stepping surrogate for macroscopic fields alternativly with direct \ac{lbm} time steps. This procedure leverages the computational efficiency of the ML model while maintaining physical fidelity through the recurrent physical bias of the discretized evolution equations, overall achieving 1 to 2 orders of magnitude in computational time savings.
\par
\begin{figure}[htbp]
    \centering
    \begin{subfigure}[c]{0.3\textwidth}
        \resizebox{0.8\linewidth}{!}{\includegraphics[]{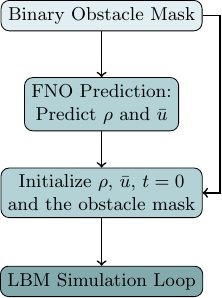}}
    \caption{Initialization Framework.}
    \label{fig:init_flowchart}
    \end{subfigure}    
    \begin{subfigure}[c]{0.69\textwidth}
        \resizebox{0.9\linewidth}{!}{\includegraphics[]{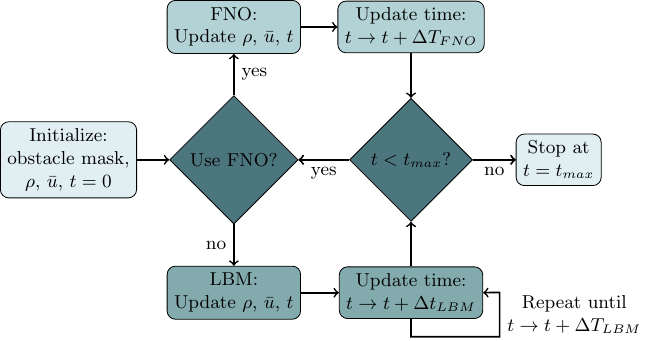}}
    \caption{Fully Hybrid FNO-LBM Framework.}
    \label{fig:hybrid_flowchart}
    \end{subfigure}
    \caption{Hybrid FNO-LBM Framework. For steady-state flows, the FNO provides an initial guess followed by the pure LBM Simulation loop depicted in \cref{fig:lbm_flowchart}. For non-steady flows, the switch criterion between LBM and FNO is defined as a fixed temporal switch based on a predefined time step.}
    \label{fig:hybrid_framework}
\end{figure}

 For steady-state problems, the FNO serves an initial guess for the macroscopic fields, which is passed to the \ac{lbm} solver. The solver then proceeds to converge to the steady solution by \ac{lbm} time marching, without further interaction with the ML model (see \cref{fig:init_flowchart}).
\par For unsteady flows, the framework alternates between FNO-based predictions and \ac{lbm} time stepping throughout the entire simulation horizon (see \cref{fig:hybrid_flowchart}). The switching frequency between \ac{fno} and \ac{lbm} is governed by a predefined temporal criterion. All \ac{fno}s employed in this work are trained to predict a fixed time increment of $\Delta T_{FNO} = 223 \cdot \Delta t_{LBM} = \Delta T$ with respect to the LBM simulation time step $\Delta t_{LBM}$, while the LBM advances the solution over multiple of its intrinsic time steps following
\begin{equation}
\Delta T_{LBM} = k \cdot \Delta T_{FNO} = k \cdot \Delta T, \quad k \in \{1,2,3\}.
\end{equation}
Based on this relation, we define the following hybrid configurations:
\begin{itemize}
\item 1:1‑Hybrid: $\Delta T$ predicted by the FNO and $\Delta T$ advanced by the LBM ($k=1$),
\item 1:2‑Hybrid: $\Delta T$ predicted by the FNO and $2 \Delta T$ advanced by the LBM ($k=2$),
\item 1:3‑Hybrid: $\Delta T$ predicted by the FNO and $3 \Delta T$ advanced by the LBM ($k=3$).
\end{itemize}
These configurations enable systematic control over the degree of ML intervention and thereby provide a tunable balance between computational speed‑up and numerical accuracy.
%


\section{Results}
\label{sec:results}
This section presents results for (i) steady‑state LBM initialization using FNO surrogates and (ii) unsteady hybrid FNO–LBM simulations. Training, simulations and rollouts are performed on an NVIDIA RTX A4000. FNO models are trained using the open‑source NVIDIA PhysicsNeMo framework \cite{nvidiamodulus2023}, with default hyperparameters unless otherwise specified. Separate networks are trained to predict the velocity $\bar{u}$ and density $\rho$ fields. Metrics are computed per field and per time step when applicable, global metrics are obtained via temporal averaging. Steady‑state results are aggregated over 20 trajectories, unsteady results over 100. Arithmetic means and standard deviations are reported for linear‑scale quantities, and geometric statistics for logarithmic ones.
\subsection{Steady-State Simulation}
\label{sec:results-steady}
\subsubsection{Porous Media Flow Setup}%
\begin{figure}
    \centering
    \includegraphics[width=0.98\linewidth]{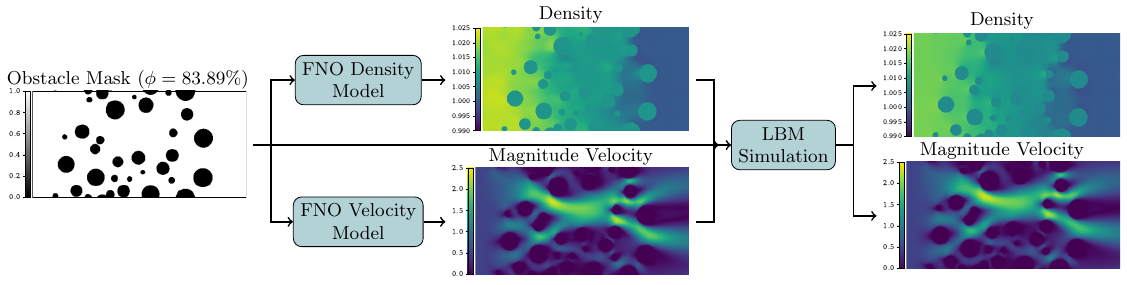}
    \caption{Exemplary porous media initialization based on the workflow introduced in \cref{fig:init_flowchart}. From left two right, the obstacle mask is created and given to the FNO models to predict density and velocity fields. The results are transfered to the LBM simulation loop that finally gives the steady-state fields (far right).}
    \label{fig:porous_workflow}
\end{figure}

We consider a two-dimensional, laminar flow through a rectangular domain ($L_{x_1}=2.0$\,m, $L_{x_2}=1.0$\,m) populated with randomly sized and distributed circular obstacles with no-slip walls. The sides are periodic, the inlet imposes $U_\infty=0.5$\SI{}{\frac{m}{s}} and the outlet a fixed pressure. Simulations use single-precision D2Q9 SRT LBM with kinematic viscosity $\nu=0.01$\SI{}{\frac{m^{2}}{s}}. Porosity spans $\phi = 0.75$--$0.90$, following 
\begin{equation}
    \phi = \frac{A_\text{pores}}{A_\text{total}}=\frac{A_\text{total}-A_\text{solid}}{A_\text{total}},
\end{equation}
and is modeled after porosity ranges investigated in previous work on porous media flows in LBM simulations \cite{espinoza2019,lbmporous2023}. A summary of the simulation parameters used for this configuration is provided in \cref{table:lbmShearSetup}. Two FNO surrogates map the binary obstacle mask to steady density and velocity fields with an illustrative workflow is shown in \cref{fig:porous_workflow}. The \ac{fno} hyperparameters for the density and velocity model as well as the dataset parameters are listed in \cref{table:porous_fno_density_hyperparameters}, \cref{table:porous_fno_velocity_hyperparameters} and \cref{table:porousdatasetparameters}.%
\subsubsection{Porous Media Flow Results}
\begin{figure}[htbp]
    \centering
    \begin{subfigure}[c]{0.49\textwidth}
        \resizebox{\linewidth}{!}{\includegraphics[]{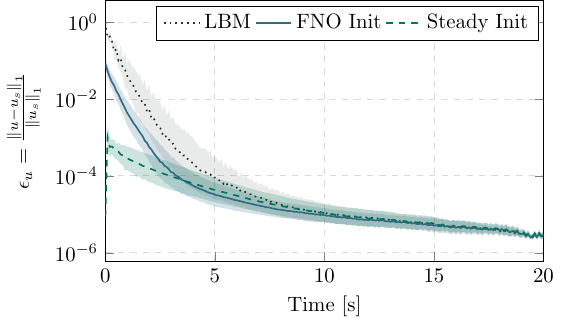}}    
        \caption{Convergence of the $u$-velocity field.}
        \label{fig:steady_u_conv}
    \end{subfigure}
    \begin{subfigure}[c]{0.49\textwidth}
        \resizebox{\linewidth}{!}{\includegraphics[]{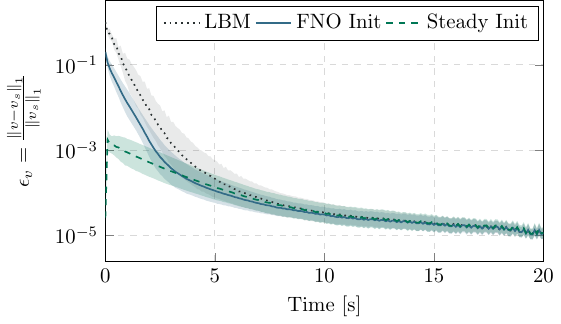}}    
        \caption{Convergence of the $v$-velocity field.}
        \label{fig:steady_v_conv}
    \end{subfigure}
    \begin{subfigure}[c]{0.49\textwidth}
        \resizebox{\linewidth}{!}{\includegraphics[]{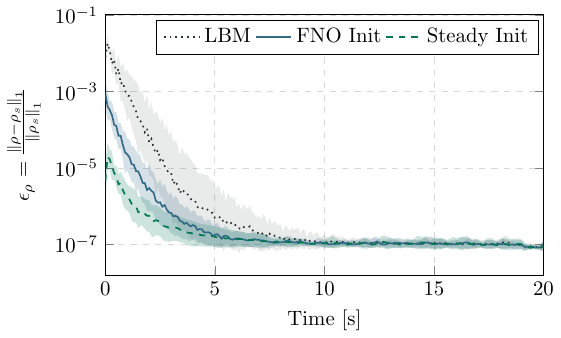}}    
        \caption{Convergence of the density field $\rho$.}
        \label{fig:steady_rho_conv}
    \end{subfigure}
    \begin{subfigure}[c]{0.49\textwidth}
        \resizebox{\linewidth}{!}{\includegraphics[]{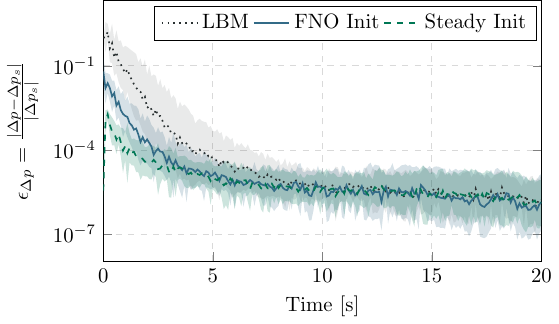}}    
        \caption{Convergence of the pressure-drop $\Delta p$.}
        \label{fig:steady_dp_conv}
    \end{subfigure}
    \caption{Convergence of the velocity components $u$ and $v$, density field $\rho$, and pressure-drop $\Delta p$, with respect to the steady-state reference solution. The steady state is computed as the average of the last 10 simulation outputs ($10 \Delta T$). The pressure-drop is defined as the difference between inflow and outflow pressures, and its error is computed relative to the steady-state pressure-drop. Shaded regions indicate standard deviation across 20 trajectories.}
    \label{fig:convergence}
\end{figure}

Convergence of the velocity components, density field, and pressure-drop relative to a steady reference solution is shown in \cref{fig:steady_u_conv,fig:steady_v_conv,fig:steady_rho_conv,fig:steady_dp_conv}. The reference is defined as the temporal average over the final $10 \Delta T$ (10 outputs) of each trajectory. Errors are reported as relative $L_1$ norms for $u$, $v$, and $\rho$, and as the relative error of the pressure-drop $\Delta p$ between inlet and outlet. Shaded bands denote the standard deviation across 20 trajectories.  
\par Initializing the LBM with FNO‑predicted fields accelerates convergence in all quantities compared with naive initialization ($\rho=1$, $u=v=0$), while all simulations ultimately reach the same steady‑state error. As summarized in \cref{tab:convergence}, speed‑ups at $\epsilon_{\mathrm{tol}}=10^{-4}$ reach $70\%$ for $\rho$ and $42.5\%$ for $\Delta p$, with positive but smaller gains of $32.7\%$ for $u$ and $11.7\%$ for $v$. Relative to initializing from previously converged fields (“Steady Init”), the FNO initialization performs comparably for $u$ and $v$, differing by only $\sim6\%$ at $\epsilon_{\mathrm{tol}}=10^{-4}$. For $\rho$, the FNO initialization enables the solver to reach the single‑precision limit ($\epsilon_{\mathrm{tol}}=2\times10^{-7}$) approximately 30\% faster than the naive LBM approach.
\begin{table}[htbp]
    \centering
    \caption{Geometric mean number of outputs required for convergence over 20 trajectories for velocity, density and pressure-drop at different tolerances $\epsilon_\mathrm{tol}$. Speed-up is reported for FNO initialization relative to pure LBM.}
    \label{tab:convergence}

    \sisetup{
        table-number-alignment = center,
        round-mode = places,
        round-precision = 2
    }

    \begin{tabular}{c l c c c c}
        \toprule
        $\epsilon_\mathrm{tol}$ & Quantity & LBM & FNO Init & Steady Init & Speed-up (\%) \\
        \midrule

        \multirow{4}{*}{$1\times10^{-4}$}
            & $u$        & 49 & 33 & 30 & \textbf{32.7} \\
            & $v$        & 63 & 53 & 49 & \textbf{11.7} \\
            & $\rho$       & 20 & 6  & 0  & \textbf{70.0} \\
            & $\Delta p$   & 40 & 23 & 9  & \textbf{42.5} \\
        \hline

        \multirow{3}{*}{$1\times10^{-5}$}
            & $u$        & 105 & 96 & 103 & \textbf{8.57} \\
            & $\rho$       & 28  & 14 & 4   & \textbf{50.0} \\
            & $\Delta p$   & 74  & 53 & 51  & \textbf{28.4} \\
        \hline

        \multirow{1}{*}{$1\times10^{-6}$}
            & $\rho$       & 42 & 26 & 14 & \textbf{38.1} \\
        \hline

        \multirow{1}{*}{$2\times10^{-7}$}
            & $\rho$       & 70 & 48 & 38 & \textbf{31.4} \\
        \bottomrule
    \end{tabular}

\end{table}
\par Overall, these results demonstrate that FNO‑based initialization provides a consistent and quantifiable acceleration of steady‑state LBM convergence without compromising accuracy.
\subsection{Unsteady Simulation}
\label{sec:results-hybrid}
\subsubsection{Double Shear Layer Setup}%
\begin{figure}
    \centering
    \includegraphics[width=0.98\linewidth]{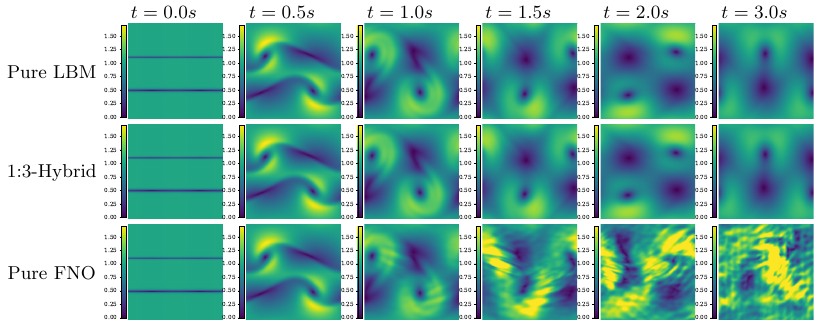}
    \caption{Exemplary rollout with the 2.6M FNO model based on the workflow introduced in \cref{fig:hybrid_flowchart}. The first row depict a pure LBM simulation, in the middle is a hybrid with one quarter of the simulation steps replaced by the FNO and the bottom row depicts the pure autoregressive rollout of the 2.6M FNO. From left two right, the analytical initialization is created and given to the FNO model and LBM simulation to predict density and velocity fields for 3s simulation time.}
    \label{fig:shear_workflow}
\end{figure}

We study the two-dimensional double shear layer in a periodic square domain ($L_{x_1}=L_{x_2}=1.0$\,m). The horizontal velocity $u$ initially follows a piecewise hyperbolic-tangent profile centered at randomized vertical positions, while the vertical velocity $v$ adds a small sinusoidal perturbation to trigger instability as described by
\begin{align}
u(x_1, x_2) &= U \cdot
\begin{cases}
\tanh\left( A_1 \left( x'_2 - x_2 \right) \right), & y \geq 0.5 \\
\tanh\left( A_2 \left( x_2 - x''_2 \right) \right), & y < 0.5,
\end{cases} \\
v(x_1, x_2) &= 0.1 \cdot \sin\left( 2\pi \left( x_1 + \frac{1}{4} \right) \right).
\end{align}
Setting $U=1.0$\,m\,s$^{-1}$ and $\nu=0.001$\,m$^2$\,s$^{-1}$, Reynolds numbers range from $\approx8$ to $50$ based on the shear-layer thickness $\delta=1/A$\,m. We train two surrogate groups that each comprise a density and a velocity model: a large 11.2M-parameter FNO and a small 2.6M-parameter FNO, see \cref{table:shear_model_groups}. The large FNO remains stable in pure autoregressive rollout, whereas the small FNO diverges after several steps. Hybrid FNO--LBM coupling follows the workflow and notation introduced in \cref{sec:hybrid_framework}, with the \ac{fno} replacing a fraction of LBM steps by FNO predictions at fixed temporal cadence (1:1, 1:2, 1:3). A comparison of pure LBM, pure FNO and 1:3-Hybrid rollout is shown in \cref{fig:shear_workflow}.%
\begin{table}[htbp]
\centering
\caption{Total parameters of the FNO models for the double shear prediction.}
\begin{tabular}{l c c c}
 \hline
 Group Name & Density Model & Velocity Model & Combined \\
 \hline
 11.2M FNO  & 5 621 835 & 5 621 875 & 11 243 710\\
 2.6M FNO & 151 811 & 2 409 771 & 2 561 582\\
 \hline
\end{tabular}
\label{table:shear_model_groups}
\end{table}
\subsubsection{Double Shear Layer Results}
\begin{figure}[htbp]
    \centering
    \begin{subfigure}[c]{0.30\textwidth}
        \resizebox{\linewidth}{!}{\includegraphics[]{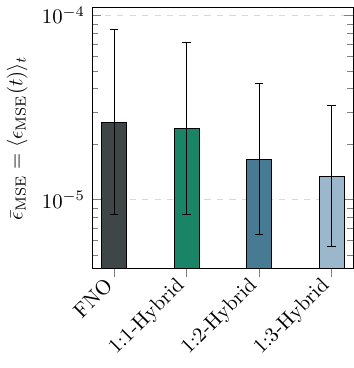}}    
        \caption{Global $\bar\epsilon_\text{MSE}$ 11.2M FNO.}
        \label{fig:hybrid_big_model_global_mse}
    \end{subfigure}
    \begin{subfigure}[c]{0.67\textwidth}
        \resizebox{\linewidth}{!}{\includegraphics[]{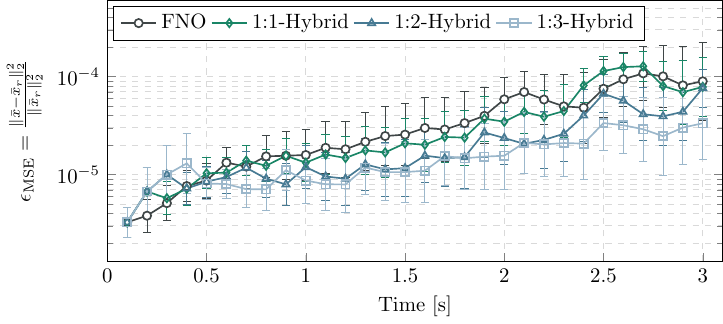}}    
        \caption{Time-dependent $\epsilon_\text{MSE}$ of the 11.2M FNO.}
        \label{fig:hybrid_big_model_mse}
    \end{subfigure}
    \begin{subfigure}[c]{0.30\textwidth}
        \resizebox{\linewidth}{!}{\includegraphics[]{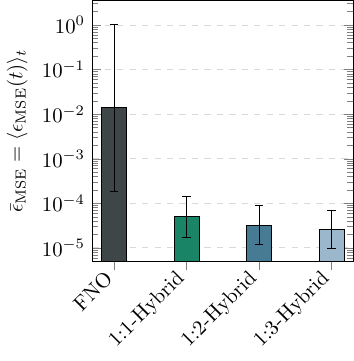}}    
        \caption{Global $\bar\epsilon_\text{MSE}$ 2.6M FNO.}
        \label{fig:hybrid_small_model_global_mse}
    \end{subfigure}
    \begin{subfigure}[c]{0.67\textwidth}
        \resizebox{\linewidth}{!}{\includegraphics[]{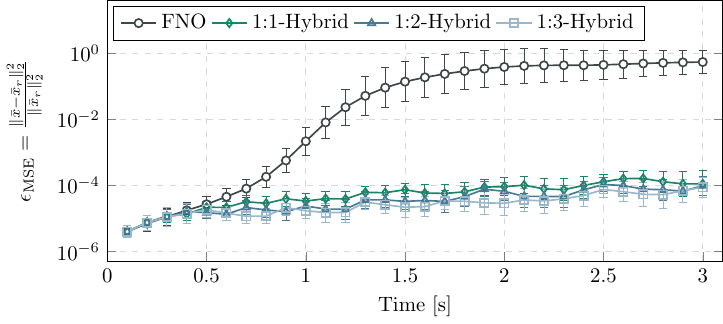}}    
        \caption{Time-dependent $\epsilon_\text{MSE}$ of the 2.6M FNO.}
        \label{fig:hybrid_small_model_mse}
    \end{subfigure}
    \caption{Global ($\bar\epsilon_\text{MSE}$) and time-dependent ($\epsilon_\text{MSE}$) MSE of the 11.2M and 2.6M FNO models in pure and hybrid rollout scenarios  with respect to the reference pure LBM simulation.}
    \label{fig:hybrid_mse}
\end{figure}

\begin{figure}[htbp]
    \centering
    \begin{subfigure}[c]{0.30\textwidth}
        \resizebox{\linewidth}{!}{\includegraphics[]{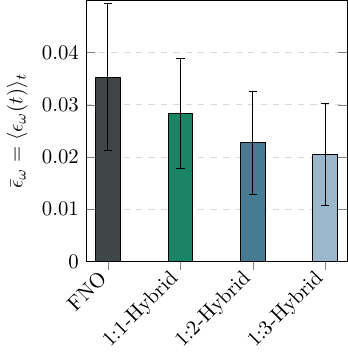}}    
        \caption{Global $\bar\epsilon_\omega$ 11.2M FNO.}
        \label{fig:hybrid_big_model_global_vorticity}
    \end{subfigure}
    \begin{subfigure}[c]{0.67\textwidth}
        \resizebox{\linewidth}{!}{\includegraphics[]{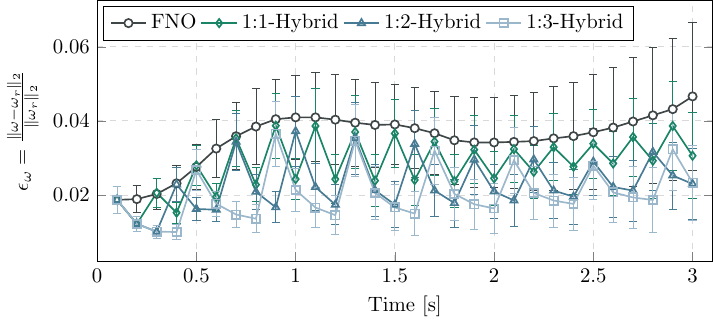}}    
        \caption{Time-dependent $\epsilon_\omega$ of the 11.2M FNO.}
        \label{fig:hybrid_big_model_vorticity}
    \end{subfigure}
    \begin{subfigure}[c]{0.30\textwidth}
        \resizebox{\linewidth}{!}{\includegraphics[]{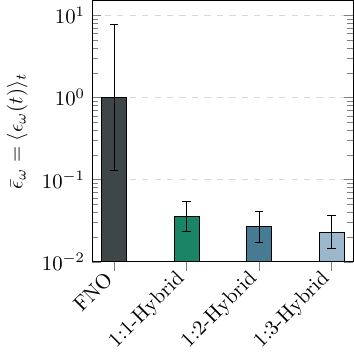}}    
        \caption{Global $\bar\epsilon_\omega$ 2.6M FNO.}
        \label{fig:hybrid_small_model_global_vorticity}
    \end{subfigure}
    \begin{subfigure}[c]{0.67\textwidth}
        \resizebox{\linewidth}{!}{\includegraphics[]{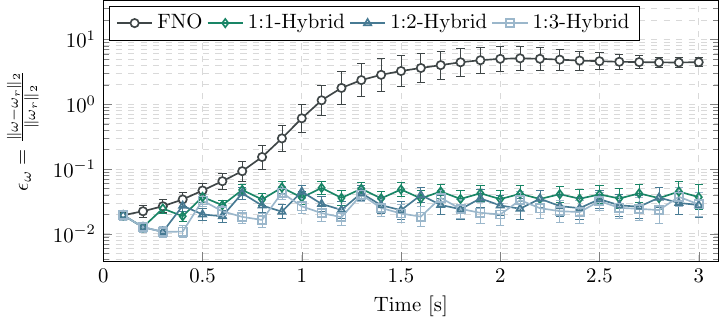}}    
        \caption{Time-dependent $\epsilon_\omega$ of the 2.6M FNO.}
        \label{fig:hybrid_small_model_vorticity}
    \end{subfigure}
    \caption{Global ($\bar\epsilon_\omega$) and time-dependent ($\epsilon_\omega$) relative $\mathcal{L}_2$ vorticity error of the 11.2M and 2.6M FNO models in pure and hybrid rollout scenarios with respect to the reference pure LBM simulation.}
    \label{fig:hybrid_vorticity}
\end{figure}

We quantify accuracy with a normalized vorticity error and a mean-squared error across all fields, both computed per time step and as temporal averages over the $30\Delta T$ horizon (\cref{fig:hybrid_mse,fig:hybrid_vorticity}). 
\par Across both surrogate sizes, the global MSE decreases monotonically with more frequent corrections (1:1$\to$1:2$\to$1:3). The effect is modest yet consistent for the 11.2M FNO, and dramatic for the 2.6M FNO, where hybrid coupling suppresses divergence and yields stable, low-error predictions over the entire rollout. Time-resolved curves corroborate these findings: hybrid variants remain uniformly bounded and below the pure-FNO baselines, with the 1:3-Hybrid providing the lowest and most consistent errors, see \cref{fig:hybrid_big_model_mse,fig:hybrid_small_model_mse}.%
\par Vorticity errors exhibit the same trend. Global $\bar{\epsilon}_\omega$ decreases with increasing correction frequency, indicating improved preservation of small-scale rotational structures. The 2.6M pure FNO shows pronounced error growth over time, whereas all hybrids remain stable and attain vorticity errors comparable to the large-model hybrid, see \cref{fig:hybrid_vorticity}.
\par \Cref{tab:hybrid_error_reduction} summarizes relative reductions in mean global errors achieved by the hybrid models compared to their pure FNO counterparts. For the 11.2M model, hybridization yields global MSE reductions of $7.9\%$, $37.1\%$, and $49.3\%$ for 1:1, 1:2, and 1:3 configurations, respectively. The corresponding vorticity reductions are $19.8\%$, $35.6\%$, and $42.1\%$. For the 2.6M model, reductions are even more pronounced with $99.6$--$99.8\%$ (MSE) and $96.5$--$97.7\%$ (vorticity), effectively transforming an otherwise divergent surrogate into one operating within the same error regime as the large-model hybrid. Collectively, these results show that hybridization improves accuracy and enforces long-horizon stability, enabling lightweight surrogates to perform competitively at a fraction of the parameter count.%
\begin{table}[htbp]
    \centering
    \caption{Relative reduction of the mean global error over 100 trajectories for hybrid FNO--LBM models. Error reduction is computed as 
    $(\bar{\epsilon}_{\mathrm{FNO}} - \bar{\epsilon}_{\mathrm{Hybrid}}) / \bar{\epsilon}_{\mathrm{FNO}}$.}
    \label{tab:hybrid_error_reduction}


    \begin{tabular}{l c c c c}
        \toprule
        Metric & Model & 1:1-Hybrid (\%) & 1:2-Hybrid (\%) & 1:3-Hybrid (\%)\\
        \midrule

        \multirow{2}{*}{$\bar{\epsilon}_\omega$}
            & 11.2M FNO  & 19.8 & 35.6 & 42.1\\
            & 2.6M FNO  & 96.5 & 97.3 & 97.7\\
        \midrule

        \multirow{2}{*}{$\bar{\epsilon}_{\mathrm{MSE}}$}
            & 11.2M FNO  & 7.9  & 37.1 & 49.3\\
            & 2.6M FNO  & 99.6 & 99.8 & 99.8\\
        \bottomrule
    \end{tabular}
\end{table}
\par Lastly, the speed-up of the hybrid approach compared to the pure LBM is investigated, see \cref{tab:hybrid_speedup}. Wall-clock runtimes are averaged over the first six trajectories with standard deviations per case of less than 0.8\% with reference to the respective mean runtime. It has to be noted, that the potential speed-up of ML models in hybrid scenarios is highly dependent on the ratio of their learned prediction time step $\Delta T$ and the time step $\Delta t$ of the numerical simulation. Both investigated models achieved nearly identical speed-ups, reaching from around 5\% to 11\%, depending on the hybrid case. The reduction of total simulations steps is identical for both.  
%
%
\begin{table}[htbp]
    \centering
    \caption{Relative reduction of simulation runtime $\epsilon_T=(T_{\mathrm{LBM}} - T_{\mathrm{Hybrid}}) / T_{\mathrm{LBM}}$ and number of simulation steps $\epsilon_N=(N_{\mathrm{LBM}} - N_{\mathrm{Hybrid}}) / N_{\mathrm{LBM}}$ for hybrid approaches compared to pure LBM simulation.}
    \label{tab:hybrid_speedup}


    \begin{tabular}{c c c c c}
        \toprule
        Metric & Model & 1:1-Hybrid (\%) & 1:2-Hybrid (\%) & 1:3-Hybrid (\%) \\
        \midrule

        \multirow{2}{*}{$\epsilon_T$}
            & 11.2M FNO  & 11.1 & 7.7 & 5.1 \\
            & 2.6M FNO  & 11.8 & 7.7 & 5.7 \\
        \midrule

        \multirow{1}{*}{$\epsilon_N$}
            & -- & 50  & 34 & 25 \\
        \bottomrule
    \end{tabular}
\end{table}
\par Taken together, these findings demonstrate that hybrid FNO--LBM models deliver substantial improvements in accuracy and long-term stability while still offering meaningful runtime savings, particularly when applied to compact surrogate architectures.
%

%
%
\section{Conclusion}
\label{sec:conclusion}
This work demonstrates that FNO surrogates can be effectively coupled with LBM solvers to accelerate both steady‑state and unsteady flow simulations while preserving physical accuracy. For steady porous media flows, FNO‑initialized LBM substantially reduces convergence time across all macroscopic fields, with the greatest improvements observed in density and pressure‑drop convergence. For unsteady double shear layer dynamics, hybrid FNO–LBM coupling consistently lowers time‑resolved and global errors, improves long‑horizon stability, and mitigates the drift typically observed in autoregressive surrogate rollouts.
\par A key outcome is the strong regularizing effect of numerical correction: hybridization enables a compact 2.6M‑parameter FNO, which is unstable in standalone inference, to operate within the same accuracy regime as an 11.2M‑parameter model, yielding reductions of up to $99.8\%$ in global MSE and $97.7\%$ in vorticity error. These results highlight hybrid coupling as an efficient mechanism for enforcing physical consistency while retaining the computational benefits of neural operators.
\par Overall, the proposed framework offers a reliable and scalable pathway for integrating machine‑learning surrogates into high‑resolution CFD workflows, and provides a foundation for future developments in accelerated, physically grounded hybrid simulation methods.
\section*{Acknowledgments}
This work was funded by the European Union under the ERC Advanced Grant: Project 101094463 — GENUFASD

\appendix
\section{Pourous Media Flow Setup}
\begin{table}[htbp]
\centering
\caption{Porous Media Flow Simulation Parameters}
\begin{tabular}{p{0.37\textwidth}p{0.27\textwidth}p{0.23\textwidth}}
 \hline
 Boundary Conditions & Physical Parameters & \ac{lbm} Algorithm \\
 \hline
 Inlet: Fixed Inflow  & $U_\infty=0.5$\SI{}{\frac{m}{s}}  & Single Precision \\
 Sides: Periodic &  \( \nu =0.01 \)\SI{}{\frac{m^{2}}{s}}& D2Q9 \\
 Outlet: Fixed Pressure & $\phi=0.75-0.9$ & \ac{srt} \\
 \hline
\end{tabular}
\label{table:lbmShearSetup}
\end{table}
\begin{table}[htbp]
\centering
\caption{Porous Media \ac{fno} Density Model Hyperparameters}
\begin{tabular}{p{0.22\textwidth}p{0.3\textwidth}p{0.38\textwidth}}
 \hline
 Decoder & FNO & Scheduler \\
 \hline
 Out Features: 1 & In Channels: 1 &  Learning Rate: 1.e-2\\
 Layers: 3 & Latent Channels: 16 & Decay Rate: 0.85\\
 Layer Size: 32 & Layers: 3 & Decay Frequency: 10.Epoch\\
                & Fourier Modes: 6 & Type: LambdaLR \\
                & Padding: 9 & Optimizer AdamW\\
 \hline
 \multicolumn{3}{l}{Total Model Parameters: 114 273}  \\
 \hline
\end{tabular}
\label{table:porous_fno_density_hyperparameters}
\end{table}
\begin{table}[htbp]
\centering
\caption{Porous Media \ac{fno} Velocity Model Hyperparameters}
\begin{tabular}{p{0.22\textwidth}p{0.3\textwidth}p{0.38\textwidth}}
 \hline
 Decoder & FNO & Scheduler \\
 \hline
 Out Features: 2 & In Channels: 1 &  Learning Rate: 1.e-2\\
 Layers: 3 & Latent Channels: 16 & Decay Rate: 0.85\\
 Layer Size: 64 & Layers: 6 & Decay Frequency: 10.Epoch\\
                & Fourier Modes: 12 & Type: LambdaLR \\
                & Padding: 9 & Optimizer AdamW\\
 \hline
 \multicolumn{3}{l}{Total Model Parameters: 896 082}  \\
 \hline
\end{tabular}
\label{table:porous_fno_velocity_hyperparameters}
\end{table}
\begin{table}[htbp]
\centering
\caption{Porous Media Dataset Parameters}
\begin{tabular}{p{0.33\textwidth}p{0.35\textwidth}p{0.22\textwidth}}
 \hline
 Dataset & Sample Size & Resolution \\
 \hline
 Training Data & 1700 &  512x256\\
 Validation Data & 200 &  512x256\\
 Inference & 20 Trajectories &  512x256\\
 \hline
\end{tabular}
\label{table:porousdatasetparameters}
\end{table}
\section{Double Shear Layer Flow Setup}
\begin{table}[htbp]
\centering
\caption{Double Shear \ac{fno} Density Models Hyperparameters}
\begin{tabular}{p{0.22\textwidth}p{0.3\textwidth}p{0.38\textwidth}}
 \hline
 Decoder & FNO & Scheduler \\
 \hline
 Out Features: 1 & In Channels: 1 &  Learning Rate: 1.e-2\\
 Layers: 1 & Latent Channels: 14 & Decay Rate: 0.85\\
 Layer Size: 32 & Layers: 3,7 & Decay Frequency: 8.Epoch\\
                & Fourier Modes: 8,32 & Type: LambdaLR \\
                & Padding: 9 & Optimizer AdamW\\
 \hline
 \multicolumn{3}{l}{Total Model Parameters (3 Layers, 8 Fourier Modes): 151 811}  \\
 \multicolumn{3}{l}{Total Model Parameters (7 Layers, 32 Fourier Modes): 5 621 835}  \\
 \hline
\end{tabular}
\label{table:shear_fno_density_hyperparameters}
\end{table}
\begin{table}[htbp]
\centering
\caption{Double Shear \ac{fno} Velocity Models Hyperparameters}
\begin{tabular}{p{0.22\textwidth}p{0.3\textwidth}p{0.38\textwidth}}
 \hline
 Decoder & FNO & Scheduler \\
 \hline
 Out Features: 2 & In Channels: 2 &  Learning Rate: 1.e-2\\
 Layers: 1 & Latent Channels: 14 & Decay Rate: 0.85\\
 Layer Size: 32 & Layers: 3,7 & Decay Frequency: 8.Epoch\\
                & Fourier Modes: 32 & Type: LambdaLR \\
                & Padding: 9 & Optimizer AdamW\\
 \hline
 \multicolumn{3}{l}{Total Model Parameters (3 Layers): 2 409 771}  \\
 \multicolumn{3}{l}{Total Model Parameters (7 Layers): 5 621 875}  \\
 \hline
\end{tabular}
\label{table:shear_fno_velcity_hyperparameters}
\end{table}
\begin{table}[htbp]
\centering
\caption{Double Shear Layer Dataset Parameters}
\begin{tabular}{p{0.33\textwidth}p{0.35\textwidth}p{0.22\textwidth}}
 \hline
 Dataset & Samples & Resolution \\
 \hline
 Training & 3000 Pairs&  128x128\\
 Validation & 1000 Pairs&  128x128\\
 Inference & 100 Trajectories &  128x128\\
 \hline
\end{tabular}
\label{table:sheardatasetparameters}
\end{table}
%


\bibliographystyle{elsarticle-num}
\bibliography{06_references}

\end{document}